\documentclass[aps,prl,showpacs,twocolumn]{revtex4}

\usepackage{amsmath,epsfig,amsfonts,epsfig,graphicx,
}

\begin{document}
\title{
Modulational instability in a layered Kerr medium: Theory and Experiment}
\author{Martin Centurion$^{1,2}$, Mason A. Porter$^{2,3}$, Ye Pu$^1$,
P. G. Kevrekidis$^4$, D. J. Frantzeskakis$^5$ and Demetri Psaltis$^{1}$}
\affiliation{ $^1$Department of Electrical Engineering,
$^2$Center for the Physics of Information,
$^3$Department of Physics, California Institute of Technology, Pasadena, CA 91125, USA \\
$^4$Department of Mathematics and Statistics, University of Massachusetts, Amherst MA 01003-4515, USA \\
$^5$ Department of Physics, University of Athens, Panepistimiopolis, Zografos, Athens 15784, Greece}

\begin{abstract}
We present the first experimental investigation of modulational instability in a layered Kerr medium.
The particularly interesting and appealing feature of our configuration, consisting of alternating glass-air layers, is the piecewise-constant nature of the material properties, which allows a theoretical linear stability analysis leading to a Kronig-Penney equation whose forbidden bands correspond to the modulationally unstable regimes.  We find very good {\it quantitative} agreement between theoretical, numerical, and experimental diagnostics of the modulational instability. Because of the periodicity in the evolution variable arising from the layered medium, there are multiple instability regions rather than just one as in the uniform medium.
\end{abstract}


\maketitle


{\it Introduction}. The modulational instability (MI) is a destabilization mechanism for plane waves.
It leads to delocalization in momentum space and, in turn, to localization in position space and the
formation of solitary-wave structures. The MI arises in
many physical contexts,
including fluid dynamics \cite{benjamin67}, nonlinear optics \cite{ostrovskii69},
plasma physics \cite{taniuti68},
Bose-Einstein condensate (BEC) physics \cite{pandim}, and so on.

The MI was originally analyzed in uniform media, mainly in the framework of the nonlinear Schr{\"o}dinger equation.  There, MI occurs for a focusing nonlinearity and long-wavelength perturbations of the pertinent plane waves \cite{benjamin67,ostrovskii69,taniuti68,pandim}.
Recently, several
experimentally relevant settings with (temporally and/or spatially) {\it nonuniform}
media have emerged.
Such research
includes the experimental observation of bright matter-wave soliton trains in BECs induced by the temporal change of the interatomic interaction from repulsive to attractive through Feshbach resonances \cite{randy}. This effective change of the nonlinearity from defocusing to focusing leads to the onset of MI and the formation of the soliton trains \cite{ourpra}.  Soliton trains can also be induced in optical settings (e.g., in birefringent
media \cite{wabnitz}). Even closer to this Letter's theme of periodic nonuniformities
is the
vast research on photonic crystals \cite{solj} and the
experimental observations of the MI in spatially periodic optical media (waveguide arrays) \cite{dnc0} and BECs confined in
optical lattices \cite{smerzi}.
%
%
%
Finally, apart from the aforementioned results pertaining to systems that are periodic in the transverse dimensions, there exist physically relevant situations for which the periodicity is in the {\it evolution variable}.  Examples were initially proposed in the context of optics through dispersion management \cite{Progress}, and have since also been studied for nonlinearity management both in optics \cite{Isaac,centurion} and BECs \cite{frm}.

In this Letter, we present the first experimental realization of MI in a setting where the nonlinearity
is periodic in the evolution variable, which here is the propagation distance. There is a fundamental
difference between such a periodic setting and a uniform one:
In the latter, there is a cutoff wavenumber above which MI is not possible.  In other words, there is a {\it single} window of unstable wavenumbers. In a periodic medium, however, {\it additional} instability windows exist for wavenumbers above the first cutoff. Our experiments were designed to demonstrate this unique feature of the layered structure.  In addition to our experiments, our investigation includes
a linear stability analysis \cite{nd,zoi}, which leads to a Hill equation (whose coefficients are periodic in the evolution variable) \cite{magnus}.  The permissible spectral bands of this equation correspond to modulationally stable wavenumbers and the forbidden bands indicate MI. The obtained experimental and analytical results are also corroborated by numerical simulations.

Our setup consists of an optical medium with periodically alternating glass and air layers.
The piecewise constant nature of the material coefficients
leads to a linear stability condition (for plane waves) along the lines of the
Kronig-Penney model of solid state physics \cite{kittel}
(generalizations of which with spatially periodic nonlinearity have been
considered in \cite{hennig}).
This allows us to compute the MI bands analytically and to compare
the experimental findings
with the theoretical
predictions.

\begin{figure}[tbp]
\centering \includegraphics[width=8.0cm]{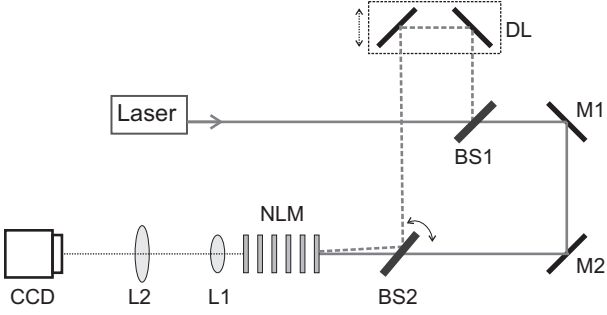}
\caption{Experimental setup.  BS1 and BS2 are beam splitters, DL is a variable delay line, M1 and M2 are mirrors, NLM is the layered nonlinear medium, and L1 and L2 are lenses.}
\label{setup}
\end{figure}

{\it Experimental Setup}. In our experiments
(see Fig.~\ref{setup}),
an amplified Titanium:Sapphire laser is used to generate 150-femtosecond pulses with an energy of 2 mJ at a wavelength of $\lambda=800$ nm. The beam profile is approximately Gaussian with a full-width at half-maximum of 1.5 mm. The laser pulses are split into a pump and a reference using a beam splitter (BS1), with most of the energy in the pump pulse. After synchronization with a variable delay line (DL), the two pulses are recombined at a second beam splitter (BS2) and sent to the periodic nonlinear medium (NLM).
The reference introduces a sinusoidal modulation in the intensity (i.e., an interference pattern),
with the period determined by the relative angle between the two beams. The angle of the reference
is carefully tuned by rotating BS2 so that the two beams overlap while propagating through the NLM at
adjustable angles. The NLM consists of six 1 mm thick quartz slides separated by air gaps.
The glass slides have an anti-reflection coating to minimize the loss (the reflection from each interface is 1\%).  The loss due to back-reflections from the slides is included in our numerical simulations below, and the effect of double reflections is negligible. In our experiments, we used structures with air gaps of 2.1 mm and 3.1 mm.  The intensity pattern after the NLM (at the output face of the last quartz slide) is imaged on a CCD camera (Pulnix TM-7EX) using two lenses (L1 and L2) in a 4-F configuration, with a magnification of $M = 8$.  The CCD camera captures the central region (0.6 mm $\times$ 0.8 mm) of the beam.

The intensity pattern at the output of the NLM is recorded both for a high pump intensity
($I_{P1}=1.3\times10^{11}$ W/cm$^2$) and a low pump intensity ($I_{P2}=9\times10^{8}$ W/cm$^2$).
In both cases, the intensity of the reference beam is 1\% of that
of the pump. We measure the effect of the nonlinearity by comparing the output
for high versus low intensity.  In the latter case, the propagation is
essentially linear.  If the spatial frequency of the modulation lies inside
the instability window, the amplitude of the reference wave increases at the expense of the pump.


{\it Theoretical Setup}. Our theoretical model for the beam propagation incorporates the dominant
dispersive and Kerr effects in a nonlinear Schr{\"o}dinger equation,
\begin{align}
i\frac{\partial u}{\partial \zeta} &= -\frac{1}{2}{\nabla}^2_\perp u - |u|^2u \,, \,\,\,\, 0 < \zeta < \tilde{l}\,\,\, \mbox{(glass)}\,,
\notag \\
i\frac{\partial u}{\partial \zeta} &= -\frac{1}{2}\frac{n_0^{(1)}}{n_0^{(2)}}{\nabla}^2_\perp u - \frac{n_2^{(2)}}{n_2^{(1)}}|u|^2u \,, \,\,\, \tilde{l} < \zeta < \tilde{L}\,\,\,\, \mbox{(air)}\,,
\label{nls}
\end{align}
where space is rescaled by the wavenumber, $(\xi,\eta,\zeta)=k^{(1)}\! \times \!(x,y,z)$, and
the electric field envelope is rescaled using $u=(n_2^{(1)}/n_0^{(1)})^{1/2} E$.
The superscript $(j)$ denotes the medium, with $j=1$ for glass and $j=2$ for air.
The Kerr coefficients of glass and air are $n_2^{(1)}=3.2\times10^{-16}$ cm$^2$/W and
$n_2^{(2)}=3.2\times10^{-19}$ cm$^2$/W, respectively. Additionally, $n_0^{(1)}=1.5$ and $n_0^{(2)}=1$.
The above setting (incorporating the transmission losses at each slide) can be written compactly as
\begin{equation}
    i \frac{\partial u}{\partial \zeta}=-\frac{1}{2} D(\zeta) \nabla^2 u - N(\zeta) |u|^2 u - i \gamma(\zeta) u\,,
\label{nls1}
\end{equation}
where $D(\zeta)$ and $N(\zeta)$ are
piecewise constant functions in consonance with Eq.~(\ref{nls}) and $\gamma(\zeta)$ is the loss rate.
Equation~(\ref{nls1}) possesses plane wave solutions of the form,
\begin{equation}
    u_0=A_0 e^{-\int^{\zeta} \gamma(\zeta') d \zeta'} e^{i A_0^2 \int^{\zeta} N(\zeta')
    \left(e^{-2\int^{\zeta'} \gamma(\tilde{\zeta}) d\tilde{\zeta}}\right) d\zeta'}\,,
\label{nls2}
\end{equation}
where $A_0$ is the initial amplitude.
We perform a
stability analysis by inserting a Fourier-mode decomposition,
$u = u_0(\zeta) \left[1+ w(\zeta) \cos(k_{\xi} \xi) \cos(k_{\eta} \eta)\right]$ (where $w=F+i B$
is  a small perturbation) into Eq.~(\ref{nls1}).
This yields
%
\begin{equation}
    \frac{d^2 F}{d \zeta^2}=\frac{1}{D}\frac{d D}{d \zeta}\frac{d F}{d \zeta}+\left[-\frac{1}{4} \bar k^4 D^2 + N \bar k^2 D |u_0|^2\right] F\,, \label{nls3}
\end{equation}
where $\bar k^2=k_{\xi}^2+k_{\eta}^2$. While one can analyze Eq.~(\ref{nls3}) directly, the weak variation of $D(\zeta)$ can be exploited by substituting $D(\zeta)$ with its average and the losses at the interfaces can be ignored. [We have checked that this has little effect on the results from Eq.~(\ref{nls3})].
Under these additional simplifications, Eq.~(\ref{nls3}) is a
Hill equation which for the piecewise-constant nonlinearity coefficient
is the well-known Kronig-Penney model \cite{kittel}. This
can be solved analytically in both glass and air (with two integration constants for each type of region).
We match the solutions at the glass--air boundaries
and obtain matching conditions at $\zeta=\tilde{l}$ and $\zeta=\tilde{L}$.
In so doing, we employ Bloch's theorem (and the continuity of $F$ and $\frac{d F}{d\zeta}$),
according to which $F(\zeta)=e^{-i \omega \zeta} H(\zeta)$, where $H$ is a periodic function of period
$\tilde{L}$ \cite{kittel}.  This yields a homogeneous $4\times 4$ matrix equation,
whose solution gives the following equation for $\omega$:
%
\begin{eqnarray}
    \cos(\omega \tilde{L}) = &-&\frac{s_{1}^{2}+{s}_{2}^{2}}{2s_{1} {s}_{2}}\sin (s_{1}\tilde{l} )
    \sin [{s}_{2}(\tilde{L}-\tilde{l} )] \nonumber \\
    &+&\cos (s_{1}\tilde{l} )\cos [{s}_{2}(\tilde{L}-\tilde{l} )] \equiv G(\bar k)\,,
\label{nls4}
\end{eqnarray}
where $s_1^2=\bar k^2 D^{(1)} (\bar k^2 D^{(1)}/4-N^{(1)} |u_0|^2)$ and
$s_2^2=\bar k^2 D^{(2)} (\bar k^2 D^{(2)}/4-N^{(2)} |u_0|^2)$. Therefore,
$|G(\bar k)|\leq 1$ implies stability and $|G(\bar k)|>1$ leads to MI.


{\it Results}. Before discussing our results, it is necessary to point out two additional assumptions.
First, we assume in our numerical simulations that the dynamics is effectively one-dimensional (1D)
along the direction of the modulation
(i.e., we use $k_{\eta}=0$ and vary $k_{\xi}$). Accordingly, we convert the 2D interference patterns
recorded on the CCD to 1D ones by integrating along the direction orthogonal to the modulation.
Second, we assume that the modulational dynamics of the (weakly decaying) central part of the Gaussian beam of the experiment is similar to that of a plane wave with the same intensity. We tested both assumptions and confirmed them a priori through
our experimental and numerical results and a posteriori through their quantitative comparison.

\begin{figure}[tbp]
\centering \includegraphics[width=8.0cm]{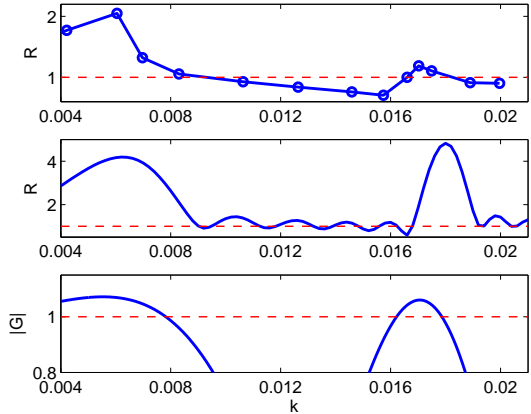} \caption{(Color online) Comparison of experimental (top), numerical (middle), and analytical (bottom) results for the 1 mm glass--2.1 mm air configuration as a function of the dimensionless wavenumber $k$. For the diagnostics $R$ and $|G|$ (defined in the text), values larger than $1$ correspond to MI.}
\label{mfig2}
\end{figure}


The input field is
$u= A_0 + \epsilon_0 \exp(ik_\xi\xi)$, where $A_0$ and $\epsilon_0$ are the amplitudes
of the pump and reference beams, respectively, and $|\epsilon_0|^2 \ll |A_0|^2$.
For linear propagation (low intensity, $I_{P2}$), the intensity pattern at the output of the NLM is approximately the same as that
at the input; it is about $|A_0|^2 +|\epsilon_0|^2 + 2A_0 \epsilon_0 \cos(k_\xi\xi)$.
For the nonlinear case (high intensity, $I_{P1}$), higher harmonics are generated and the intensity is
about $|A_1|^2 + 2 A_1 \epsilon_1 \cos(k_\xi\xi)+ 2A_1 \epsilon_2 \cos(2k_\xi\xi)+\cdots$,
where $A_1$ and $\epsilon_{n}$ ($n=1, 2, \cdots$) are, respectively, the amplitudes of the pump beam
and the $n$th harmonic at the output of the NLM. The Fourier transform (FT) of this latter intensity is
$|A_1|^2 \delta (f_\xi) + A_1 \epsilon_1 \delta(f_\xi - k_\xi/ 2\pi) + A_1 \epsilon_1 \delta(f_\xi + k_\xi/ 2\pi)
+ A_1 \epsilon_2 \delta(f_\xi - k_\xi/\pi) + A_1 \epsilon_2 \delta(f_\xi + k_\xi/\pi) + \cdots$.
The ratio of the first and zeroth order peaks in the FT is approximately equal to the ratio of the
amplitudes of the reference and pump waves: $r_0 = \epsilon_0/A_0$ and $r_1 = \epsilon_1/A_1$.
(For the experimental value of $\epsilon_0 = A_0/10$, the error introduced by this approximation is roughly $1\%$.)
The value of $r_1$ increases with propagation distance as the amplitude of the reference increases.
In the linear case, $r_0$ is constant.
We use the ratio $R = r_1/r_0$ as a diagnostic measure for both our experimental and numerical results
(so that $R > 1$ indicates growth of the perturbation). This measurement is equivalent to the ratio
$r_1(\zeta=\bar\zeta)/r_1(\zeta=0)$ (where $\bar\zeta$ is the scaled NLM length)
but is more robust experimentally.
In the numerical simulations, the peaks in the FT are sharp, whereas they are broader
in the experiments. Thus, when computing $R$ from the experiments, we used the area under the peaks instead of the peak value.

\begin{figure}[tbp]
\centering \includegraphics[width=8.0cm,height = 2.5cm]{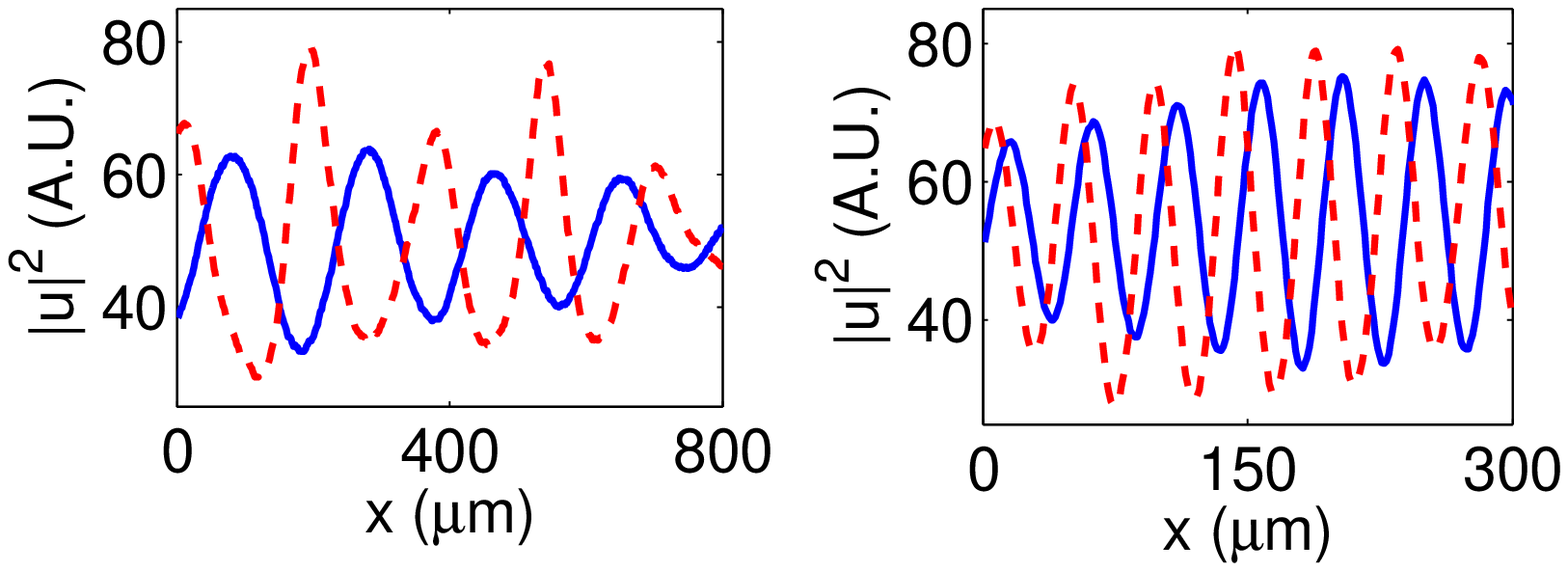}
\centering \includegraphics[width=7.8cm,height = 5.0cm]{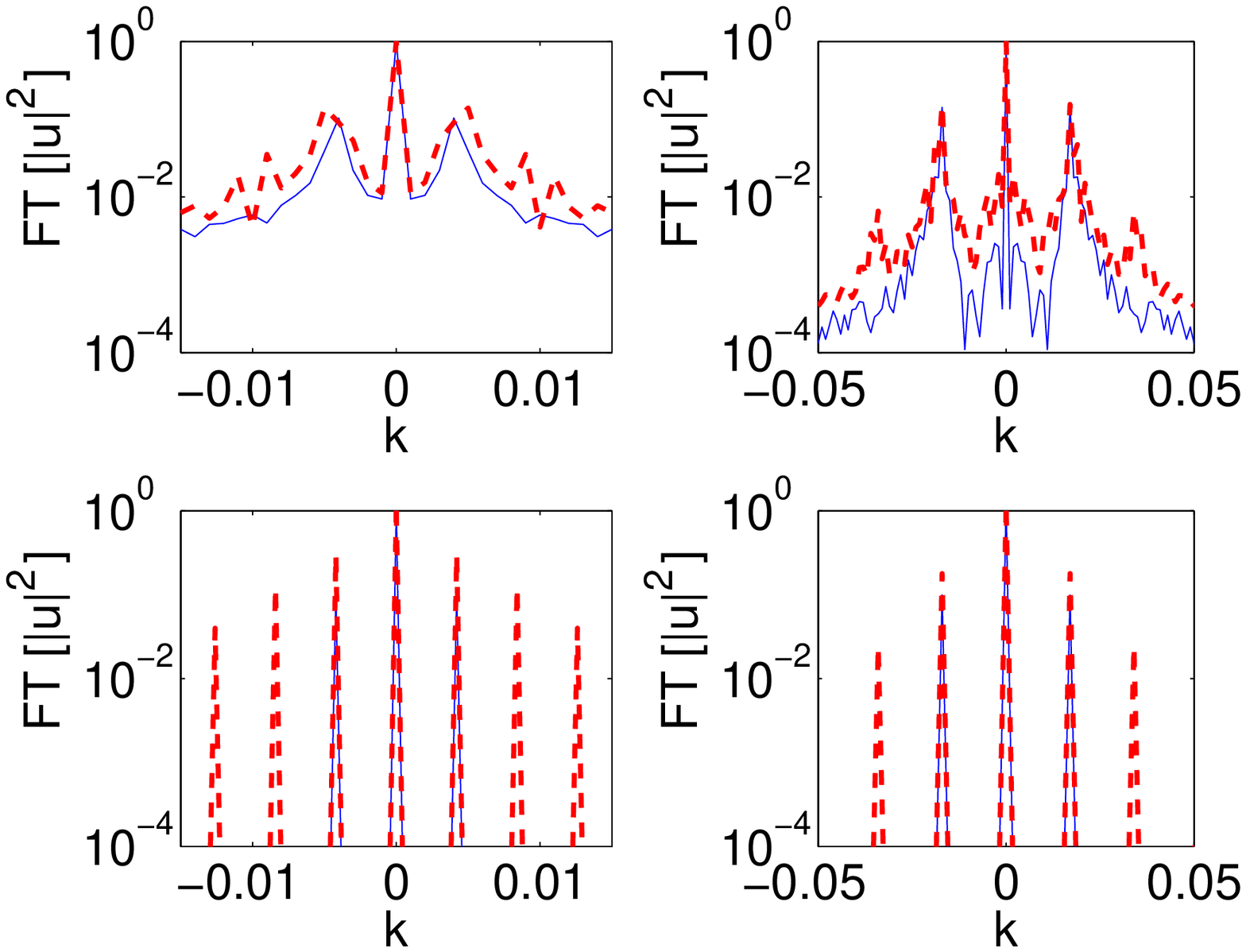}
\caption{(Color online) Experimental normalized 1D intensity patterns (top) and Fourier
spectra (middle) and numerical Fourier spectra (bottom) at the end of propagation ($z=16.5$ mm)
of the layered structure with six 1 mm glass slides, each pair of which sandwiches 2.1 mm of air.
We depict the wavenumbers $k=0.0042$ (first instability band; left panels) and $k=0.017$ (second band; right panels).  The dashed (solid) curves are for high (low) intensity $I_{P1}$ $(I_{P2})$.
}
\label{mfig3}
\end{figure}

Figure \ref{mfig2} shows the ratio $R(k)$ (where $k=k_\xi \eta_0^{(1)}$ is the sine of the angle between the pump and reference beams) for the structure with $2.1$ mm air spacings. There are two instability bands (quantified by $R > 1$) within the measurement range.
The appearance of the second band is a unique feature of the layered medium that originates from
the periodicity of the structure in the evolution variable. The maximum growth of the perturbation
in the first and second bands appear at $k=6.0\times 10^{-3}$ and $k=1.70\times 10^{-2}$,
with values of $R=2.05$ and $R=1.19$, respectively. The increase in the modulation is clearly
visible in the 1D intensity patterns (see Fig.~\ref{mfig3}).
The position of the instability bands
is in very good agreement with both numerical and theoretical ($|G|>1$) predictions.
However, the simulation typically shows a stronger instability than the experiment.
This results from the latter's 3D nature, which
is not captured in the simulation.
In the experiment, the spatial and temporal overlap of the two beams decreases with
increasing $k$, resulting in weakening of the higher-order peaks. The nonlinearity also has a lower aggregate strength in the experiment because of temporal dispersion. Nevertheless,
the simulations successfully achieve our primary goal of quantitatively capturing the locations of the instability windows.

\begin{figure}[tbp]
\centering \includegraphics[width=8.0cm]{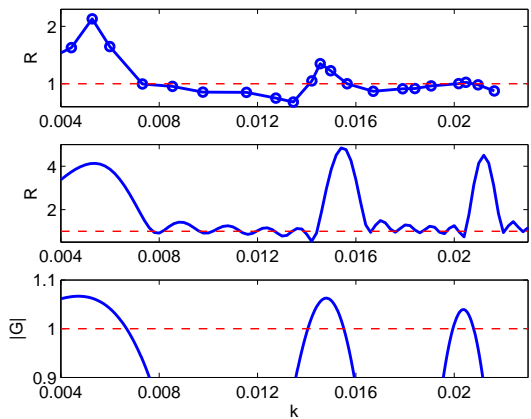}
\caption{(Color online) Same as Fig.~\ref{mfig2} but for the 1 mm glass--3.1mm air configuration.
Here there is a third MI band.}
\label{mfig4}
\end{figure}

The top panels of Fig.~\ref{mfig3} show the normalized 1D intensity pattern at the output of the NLM
for $I_{P1}$ (dashed curve) and $I_{P2}$ (solid curve). The left panels are for $k=4.2\times 10^{-3}$
(in the first instability band), and the right ones are for $k=1.70\times 10^{-2}$ (in the second band). The amplitude of the modulation increases for high intensity cases due to MI.  We have also observed in the FT of the intensity patterns (middle panels of Fig.~\ref{mfig3}) the appearance
of higher spatial harmonics of the initial modulation in the instability regions.  The first-order peaks (the ones closer to $k=0$) correspond to the modulation of the input beam and are present for both low and high intensity. The harmonics correspond to the narrowing of the peaks in the spatial interference pattern and appear only for high intensity. Strong harmonics are expected only within the instability regions.
We also observed this harmonic generation in numerical simulations (bottom panels of Fig.~\ref{mfig3}),
in good agreement with the experiments. In contrast, we did not observe such harmonics in the experiments for $k$ values in the stable regions ($R < 1$ in Fig.~2), again in agreement with theory.

Figure \ref{mfig4} shows the instability windows
for a structure with a different periodicity (with 3.1 mm air between each pair of 1 mm glass slides).
The instability bands shift towards lower $k$ and, as expected by Bloch theory, the longer spatial period in the structure results in a smaller spacing between the instability bands in Fourier space.
The peaks of the first two bands are at $k=5.3\times 10^{-3}$ and $k=1.46\times 10^{-2}$, respectively,
and a third band appears around $k=2.05\times 10^{-2}$.  Once again, we obtain good agreement between experiment, numerics, and theory.  It is not straightforward to give an intuitive explanation of the
precise location/width of the stability bands or instability zones
[beyond
employing the simple transcendental expression of Eq.~(\ref{nls4})].
However, by examining simplified cases (e.g., very narrow air gaps between wide glass
slides or vice versa) \cite{kittel}, one can extract useful information,
such as the decreasing width of the MI zones for increasing zone index
(which can be seen, e.g., in Figs.~\ref{mfig2} and \ref{mfig4}).

{\it Conclusions}. We
provided
the first experimental realization of modulational instability (MI)
in a medium periodic in the evolution variable.
The linear stability analysis of the pertinent plane waves led to an effective Kronig-Penney model,
which was used to describe
the instability bands {\it quantitatively},
providing a direct association of the MI bands with the latter's
forbidden energy zones.  One of the hallmarks of the periodic medium is
the presence of additional MI bands as opposed to the single band that
occurs in uniform media.
We found very good agreement between our theoretical predictions for the
modulationally unstable bands and those obtained experimentally and
numerically.
We also observed higher spatial harmonics for modulationally unstable beams (another characteristic trait of MI).
Many interesting extensions are possible,
including
the study of solitary waves that result from MI in layered Kerr media.

{\it Acknowledgements}. We acknowledge support from the DARPA Center for Optofluidic Integration (D.P.),
the Caltech Information Science and Technology initiative (MC, MAP), and NSF-DMS and CAREER (P.G.K.).

\end{document}